# ELECTRON SOURCE BASED ON EMERGENCE OF SELF-INJECTED ELECTRON BUNCH AT PLASMA WAKEFIELD EXCITATION BY A TW LASER PULSE

*D. S. Bondar[1], V. I. Maslov[1,2], and I. N. Onishchenko[1]*

[1]*National Science Center "Kharkiv Institute of Physics and Technology", Kharkiv, Ukraine*
[2]*Deutsches Elektronen-Synchrotron DESY, Hamburg, Germany*
*E-mail: bondar.ds@yahoo.com*

Wakefield acceleration methods are known due to some their advantages. The main of them is the high accelerating gradient up to several teravolts per meter. In the paper another important advantage is concluded to the possibility of using a wakefield accelerator as a source of electrons by means of obtaining self-injected bunches and their acceleration. The result is the simulation of the process of plasma wakefield excitation by a laser pulse with an energy of tens of mJ and a power of 1-2 TW for obtaining the promising electron source. Homogeneous and Gaussian plasma profiles were investigated and compared to increase the energy of the self-injected bunches. The laser parameters were taken that corresponded to the parameters of the laser setup in the Institute of Plasma Electronics and New Methods of Acceleration of the National Scientific Center "Kharkiv Institute of Physics and Technology". Based on the results of the simulation, the possibility of obtaining relativistic self-injected bunches that can be used for further laser acceleration experiments, including dielectric laser acceleration, was demonstrated.

## INTRODUCTION

Wakefield acceleration methods have attracted attention for their promise since the beginning of their research. This is especially true for wakefield acceleration in high-density plasma. It was researched that wakefield acceleration, driven by high-intensity lasers, can effectively accelerate electrons to high energies, with significant advancements demonstrating GeV and over 100 MeV acceleration.

It was shown the potential for developing future accelerators using this method, supported by advancements in laser technology that enable the generation of extreme intensities. Additionally, simulations have furthered the understanding of wakefield dynamics, particularly in high-frequency laser pulse applications. The advantages of wake acceleration in the creation of compact high-gradient accelerators are demonstrated [1-10].

Research into self-injected bunches formed in wake acceleration is constantly evolving. To clarify and improve the parameters of the bunches, both simulation and experiments are completed. In [11-13] it was investigated the issue of the formation and dynamics of self-injected bunches in a plasma, the density of which is approximately equal to the density of electrons in metals. It was shown that self-injected bunches formed during wake acceleration can be easily accelerated by the field, longitudinal momentum values reaching hundreds of $m_e c$ (ultrarelativistic bunches). At the same time, instabilities and methods for their suppression are demonstrated.

One of the most effective controlling self-injection process and formation of self-injected bunches can be provided by using various plasma profiles. In particular authors of [14] demonstrate control of charge, bunch length, and degree of bunch separation in a sequence using plasma density profiling. In [15] by using longitudinal plasma profiling, the position of the injection points is controlled and the electron energy is fine-tuned. In [16] it was shown that by varying the longitudinal profile of the plasma, it is possible to control the phase velocity of the wake wave. In [17, 18], the authors considered longitudinally increasing plasma density to keep the self-injected bunch in the accelerating phase of the excited wakefield. In this case due to the contraction of the wakefield bubble the velocity of its back side with maximal plasma density and excited wakefield is increasing more quickly than the laser pulse group velocity and the velocity of front side of wakefield bubble is decreasing. In result the staying time of the self-injected bunch in the high-density region with maximal accelerating wakefield should be increased. Moreover, the motion of the self-injected bunch can be synchronized with motion of bubble back side to provide radical increase of the self-injected bunch energy. The maximum of the Gaussian distribution corresponded to a 4-fold increase on the density scale (Fig. 1).

In [19] a method for controlling the parameters of self-injected bunches using laser driver profiling is discussed separately.

The resulting self-injected bunches and fields are often difficult to study due to the nonlinearity and self-consistency of the processes, which complicates the diagnosis of some processes separately from others. Nevertheless, similar studies are being carried out, as for self-injected bunches, certain concepts from external injection investigations are applicable. In [20, 21] the authors study the transformation ratio.

In particular, it is shown that to increase the transformation ratio it is necessary to use detuning of the plasma and resonant frequencies. In [22-25], the authors studied the focusing of bunches of charged particles in plasma, including in the wakefield process. In particular, the processes of plateau formation in the region of witness bunches were studied in detail. Methods for forming the best configuration of acceleration and focusing fields to obtain stable focused bunches and their acceleration are investigated.

After thorough studies of the wake process and methods of formation, acceleration, stabilization of bunches, as well as after considering various plasma density profiles, the next step should be to study the practical application of the data obtained.

One of the significant problems of laser wake acceleration is the need for powerful lasers that are guaranteed to allow self-injection and acceleration in experiment (or simulation). The power scale of such lasers is 77 TW [26], 100 TW [27], 75 TW [28].

Nevertheless, for laboratories and applied tasks it is often necessary to use lasers of significantly lower power. For this reason, the use of sub-TW lasers has been developed. It has been shown that lasers with power 0.25 TW [29], 0.6 TW [30], 0.5 TW [31] can excite wakefield to obtain self-injected bunches. The problem with sub-TW lasers is often small longitudinal momentum of self-injected bunches. In addition, often, when we talk about such lasers as real laboratory installations, both sub-TW and lasers with a power of 1-2 TW, it is not always possible to achieve the required intensity and pulse duration. And self-injected bunches have a momentum of only 1-5 $m_e c$ [29, 30]. In addition, an extremely small charge of the bunch (femtocoulomb-scale) is observed [29].

In this paper by using 2.5D numerical simulation by the fully relativistic code OSIRIS [32] considered the wakefield excitation by 1-2 TW laser pulse, the formation of a self-injected bunch and the use of inhomogeneous plasma profile to improve bunch parameters.

## STATEMENT OF THE PROBLEM

This article examines the physical processes that occur in an electron source operating on the principle of wakefield acceleration induced by a laser pulse.

The tasks of the study were to demonstrate, for considered parameters, the formation of self-injection bunch in the case of a homogeneous plasma profile and an increase in longitudinal momentum with staying of the bunch in the acceleration phase when using an inhomogeneous plasma density gradient (Gaussian density profile). The Gaussian profile was chosen due to its promising practical implementation [29]. Table 1 presents the main parameters of the laser pulse. The laser injection frequency was 10 Hz. For this reason, simulation of the injection of one beam was considered, since the simulation time is (0.417 μs << 0.1 s) significantly smaller than time between pulses, due to which the plasma will have time to return to its original state.

| Parameter | Value |
|---|---|
| Laser wavelength $\lambda_l$ | 800 nm |
| Laser radius $r_l$ at focusing point | 2.2 μm |
| Laser power $P_l$ | 1, 2 TW |
| Laser duration $\tau_l$ | 20.4 fs |
| Laser pulse energy $\varepsilon_l$ | 20.4 mJ, 40.8 mJ |
| Intensity distribution | Gaussian |
| Standard Deviation σ | 1.527 μm |

*Table 1. Laser setup and laser pulse parameters.*

These parameters corresponded to the laser setup of the Institute of Plasma Electronics and New Acceleration Methods of the National Scientific Center "Kharkiv Institute of Physics and Technology" [33, 34]. Table 2 presents the main plasma, system and simulation parameters. More detailed information about the code, as well as visualization of the pulse distribution relevant to the code, can be found in [35].

Normalization of units in simulation and visualization of results: all units of length in simulation are normalized to $c/\omega_{pe}$; all time units to $1/\omega_{pe}$; electric field and forces normalized a standard way (on breaking field) $E` = (ce/\omega_{pe})/m_e c^2 \cdot E$ (Table 2); momentum normalized on $m_e c$; energy normalized to $m_e c^2$;

The authors have previously shown the feasibility, from the point of view of keeping a self-injected bunch in the acceleration phase, of using a linearly increasing plasma density profile, provided that the density increases starting from the moment the bunch is formed in the high-density region in the tail of the wake bubble. The maximum laser intensity in the simulation coincides with the focal point (Fig. 1).

| Parameter | Value |
|---|---|
| Plasma density $n_{e0}$ | $1.74 \cdot 10^{19}$ cm$^{-3}$ |
| Spatial unit $c/\omega_{pe}$ | 1.27 μm |
| Time unit $1/\omega_{pe}$ | 4.25 fs |
| Electric field arb. un. E` | $0.401 \cdot E$ TV/m |
| Simulation windows length ($x_1$) | 80 $c/\omega_{pe}$=101.6 μm |
| Simulation windows length ($x_2$) | 20 $c/\omega_{pe}$=25.4 μm |
| Total simulation time | 417 fs |

*Table 2. Main plasma, system and simulation parameters.*

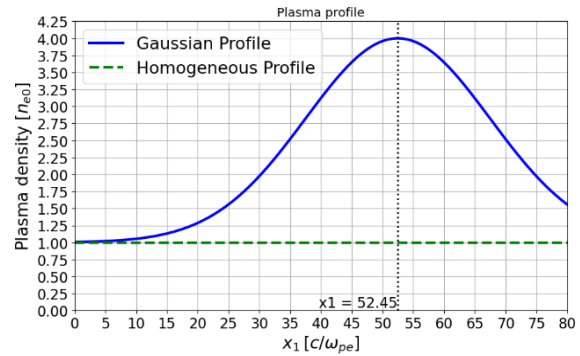

*Fig. 1. Plasma electron density longitudinal profile $n_e(x)$. The position of the laser pulse is shown.*

## RESULTS OF SIMULATION

**Homogeneous plasma profile.** It is well known that self-injected bunches are formed exclusively in the non-linear mode of the wakefield excitation in the plasma. Based on the results of the investigation, the most stable cases of self-injected bunches are presented. Based on this, the optimal laser power was used. In the case of a homogeneous plasma, a laser power of 1 TW and 2 TW were considered.

First value (**1 TW**) was chosen to combine the factors of the task of using low laser power for simulation and the acceptability of this power value to ensure self-injection. The intensity in this case was $I_1^{hm}=6.58·10^{18}$ W/cm$^2$. In Fig. 2 shown the initial moment of injection of the first bunch. One-dimensional graphs of the dependence on $x_1$ are plotted at $x_2=10$ (system axis). The simulation is performed in Cartesian coordinates in order to evaluate the full dynamics, without simplifications. Figure 3 shows the formation of the second self-injected bunch. Its formation is observed 35.7 fs later than the first bunch. Further, as it approaches the middle of the wake bubble, the value of the acceleration field will fall, after which the bunch will enter the deceleration phase.

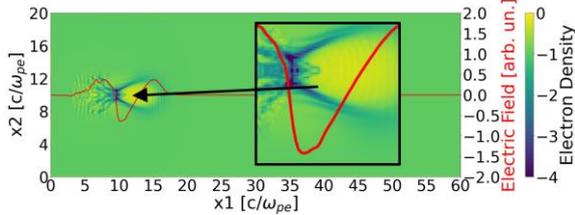

*Fig. 2. Start of acceleration (1st bunch), t=35.7 fs. Density graph $n_e(x_1, x_2)$, longitudinal acceleration field $E_x(x_1)$. Homogeneous plasma (1 TW).*

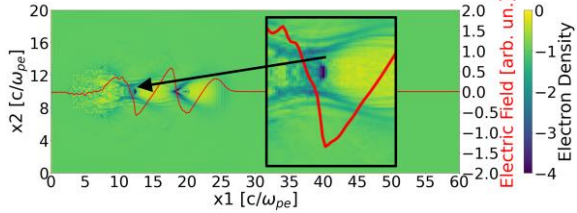

*Fig. 3. Start of self-injection (2nd bunch), t=71.4 fs. Density graph $n_e(x_1, x_2)$, longitudinal acceleration field $E_x(x_1)$. Homogeneous plasma (1 TW).*

Upon reaching the limit of the bunch being in the acceleration phase, it is supposed to be removed from the system for further operation. In Fig. 4, 5 show the final stages of self-injected bunches acceleration. Table 3 presents the parameters of self-injected bunches.

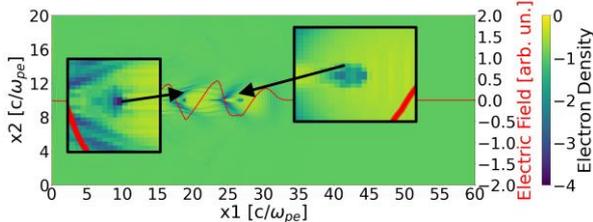

*Fig. 4. End of acceleration (1st bunch), t=95.2 fs. Density graph $n_e(x_1, x_2)$, longitudinal acceleration field $E_x(x_1)$. Homogeneous plasma (1 TW).*

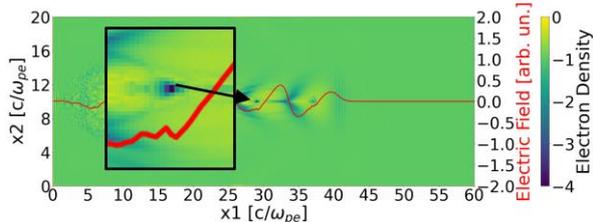

*Fig. 5. End of acceleration (2nd bunch), t=130.9 fs. Density graph $n_e(x_1, x_2)$, longitudinal acceleration field $E_x(x_1)$. Homogeneous plasma (1 TW).*

In Fig. 6 shows the longitudinal momentum of self-injected bunches at the final stages of their stay in the acceleration field phases.

Background plasma electrons are excluded from the results. Table 3 presents the results of the study of bunch parameters. Fig. 7 shows a self-injected bunch at a laser power of 2 TW in a homogeneous plasma.

| Laser pulse power | 1 TW | |
|---|---|---|
| Bunch parameter | Value | |
| | 1st bunch | 2nd bunch |
| | Final time moment | |
| | 95.2 fs | 130.9 fs |
| Length | 1.186 μm | 1.177 μm |
| Diameter | 0.802 μm | 0.7675 μm |
| Density (peak) | 2.4 $n_{e0}$ | 4.4 $n_{e0}$ |
| Charge | 4·10$^3$ fC | 6.7·10$^3$ fC |
| Average longit. momentum $p_1$ | 8.39 $m_e c$ | 6.5 $m_e c$ |
| Energy (peak) | 8.46 $m_e c^2$ 4.32 MeV | 6.6 $m_e c^2$ 3.37 MeV |
| Maximum acceleration field in the bunch region | 152 GV/m | 80 GV/m |

*Table 3. Parameters of self-injected bunches (homogeneous density distribution).*

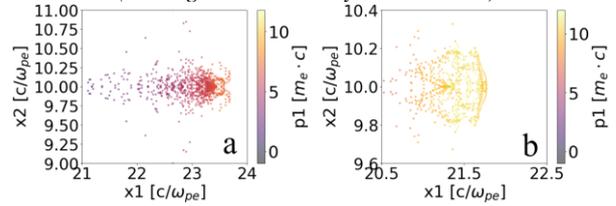

*Fig. 6. Longitudinal momentum of the bunch $p_1(x_1, x_2)$. (a) t=95.2 fs, first bunch; (b) t=130.9 fs, second bunch. Laser power = 1 TW. Homogeneous plasma.*

Fig. 7 shows a self-injected bunch at a laser power of 2 TW in a homogeneous plasma.

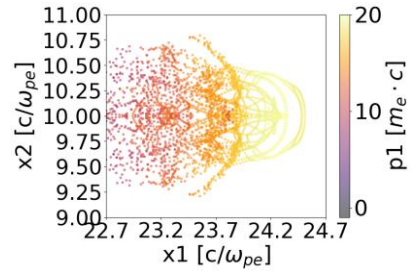

*Fig. 7. Longitudinal momentum of the bunch $p_1(x_1, x_2)$. t=107.1 fs. Laser power = 2 TW. Homogeneous plasma.*

The first bunch has a shorter longitudinal length (1.186 μm instead of 1.309 μm for the second) and slight lower maximum particle energy (4.32 MeV instead of 3.37 MeV), which indicates more efficient electron acceleration. It is also characterized by a higher longitudinal component of the momentum (8.39 $m_e c$ instead of 6.5 $m_e c$), which indicates a higher energy of the accelerated particles. The second bunch has a higher density (4.4 $n_{e0}$ instead of 2.4 $n_{e0}$) and charge (6.7·10$^3$ fC instead of 4·10$^3$ fC), which contributes to better stability and an increase in the number of particles. The maximum accelerating field in the region of bunch formation is higher for the first (152 GV/m instead of 80 GV/m for the second), which indicates a more powerful electric field and effective acceleration of particles. The first bunch has a greater increase in energy. The second bunch has the lower peak particle kinetic energy. The first bunch is in a stronger acceleration field and has acquired more energy. Differences in the characteristics of the bunches allow choosing between them depending on technical feasibility. Both bunches, with their specific characteristics, can meet the requirements for electron source generation in a variety of applications.

The parameters of the self-injected bunch in the **homogeneous** case at a laser power of **2 TW** ($I_2^{hm}=13.16 \cdot 10^{18}$ W/cm$^2$) are presented in Table 4 in comparison with the case of inhomogeneous plasma. A clear advantage of using inhomogeneous plasma for the formation of higher-energy bunches (40.19% more), bunches with a larger longitudinal momentum (39.69%) is shown. The use of inhomogeneous plasma allows increasing the acceleration field in the region of the self-injected bunch by 41.85%. Fig. 8 shows the final stage of acceleration of a self-injected bunch in a homogeneous plasma (laser power is 2 TW).

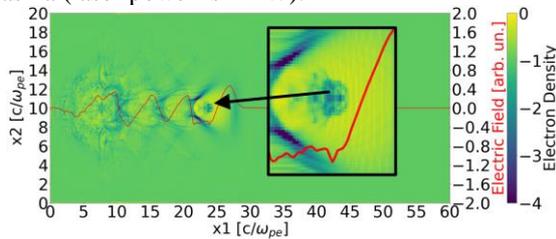

*Fig. 8. End of acceleration, t=107.1 fs. Density graph $n_e(x_1, x_2)$, longitudinal acceleration field $E_x(x_1)$. Homogeneous plasma (2 TW).*

**Inhomogeneous plasma profile.** In the case of using inhomogeneous (Gaussian) plasma (Fig. 1), one self-injected bunch is formed. In the case of an inhomogeneous plasma profile, a laser pulse power of **2 TW** is considered (1 TW is not enough for self-injection in inhomogeneous case). Power is $I_1^{inh}=13.16 \cdot 10^{18}$ W/cm$^2$. Self-injected bunch parameters and dynamics were studied. Table 4 presents the parameters of the bunch at the moment it is in the maximum acceleration field. Fig. 9 shows the time moment when self-injected bunch reaches the maximum of acceleration field.

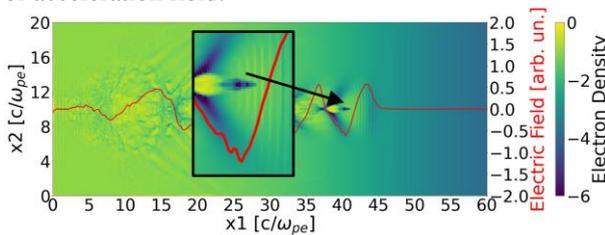

*Fig. 9. End of acceleration, t=142.8 fs. Density graph $n_e(x_1, x_2)$, longitudinal acceleration field $E_x(x_1)$. Inhomogeneous plasma.*

Nevertheless, the bunch continues its movement along the wake bubble, although already in the deceleration phase. Fig. 10 shows the shape and self-injected bunch longitudinal momentum in inhomogeneous plasma. The maximum particle energy is 11.17 MeV, and the accelerating field is 240.6 GV/m, which exceeds the parameters of bunches in a homogeneous plasma. The bunch formed in an inhomogeneous plasma shows clear advantages, including increased density, energy and a high accelerating field. As already mentioned, the comparison of the bunches parameters (Table 4) indicates obvious advantages of using an inhomogeneous plasma profile.

In addition, the data from Table 5 indicate advantages of the inhomogeneous case considered in this paper compared to previous studies by other authors [29]. The bunch has a length of 1.657 μm, a diameter of 0.828 μm and a density of 7.5 $n_{e0}$, which is significantly higher than for bunches in a homogeneous plasma. The average longitudinal component of the momentum reaches 18.19 $m_ec$, which is higher than the values observed for bunch in a homogeneous plasma. The current paper (Table 5) proposes a higher laser peak power (2 TW) and electron peak kinetic energy (11.17 MeV) compared to the work [29], which reported 0.25 TW and 0.1 MeV, respectively.

| Laser pulse power | 2 TW | |
|---|---|---|
| Bunch parameter | Value (inhomogeneous) | Value (homogeneous) |
| Time moment | 142.8 fs | 107.1 fs |
| Length | 1.657 μm | 2.159 μm |
| Diameter | 0.828 μm | 1.651 μm |
| Density (peak) | 7.5 $n_{e0}$ | 2.7 $n_{e0}$ |
| Charge | $1.9 \cdot 10^4$ fC | $3.5 \cdot 10^4$ fC |
| Average longit. momentum $p_1$ | 18.19 $m_ec$ | 10.88 $m_ec$ |
| Energy (peak) | 21.87 $m_ec^2$ <br> 11.17 MeV | 13.19 $m_ec^2$ <br> 6.74 MeV |
| Absolute energy spread ΔE | 3.09 $m_ec^2$ <br> 1.58 MeV | 2.02 $m_ec^2$ <br> 1.03 MeV |
| Relative energy spread ΔE/$E_{peak}$ | 14.14 % | 15.30 % |
| Maximum acceleration field in the bunch region | 240.6 GV/m | 139.9 GV/m |

*Table 4. Parameters of self-injected bunch In homogeneous and inhomogeneous plasma.*

| Parameter | Current paper | Z-H. He et al. [29] |
|---|---|---|
| Peak plasma density | (1,74-4.35) $\cdot 10^{19}$ cm$^{-3}$ | (0.5−2) $\cdot 10^{19}$ cm$^{-3}$ |
| Laser wavelength | 800 nm | 800 nm |
| Laser peak power | 2 TW | 0.25 TW |
| Laser pulse energy | 40.8 mJ | 8 mJ |
| Focal spot FWHM | 4.4 μm | 2.5 μm |
| Laser pulse FWHM duration | 20.4 fs | 32 fs |
| Electron peak kinetic energy $E_{peak}$ | 11.17 MeV | 0.1 MeV |
| Absolute energy spread ΔE | 1.58 MeV | 0.02 MeV |
| Relative energy spread ΔE/$E_{peak}$ | 14.14 % | 20 % |
| Longitudinal momentum $p_1$ | 18.19 $m_ec$ | 2 $m_ec$ |
| Bunch charge | $1.9 \cdot 10^4$ fC | 10 fC |

*Table 5. Comparison of current results (partially) and previous investigations for inhomogeneous plasma.*

The laser pulse energy is also substantially greater (40.8 mJ instead of 8 mJ). It was shown the formation of more intense bunches with higher charge ($1.9 \cdot 10^4$ fC instead of 10 fC). These results confirm enhanced beam characteristics under wakefield excitation in the current paper.

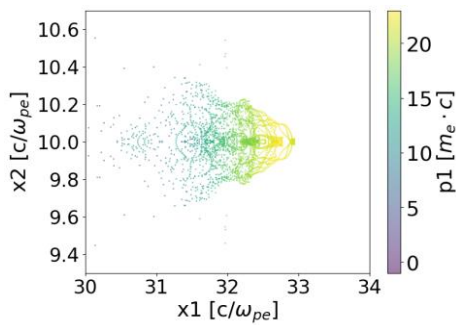

*Fig. 10. End of self-injection, t=142.8 fs. Longitudinal momentum of the bunch $p_1(x_1, x_2)$. Inhomogeneous plasma.*

## CONCLUSIONS

In this paper, using numerical simulation, it was studied the excitation of the wakefield in a plasma by a laser pulse, the parameters of which corresponded to the parameters of a laser setup operated at the NSC KIPT. It was shown for relatively high-density plasma the relevance of using this setup, obtaining parameters of bunches acceptable for further research and a high rate of their acceleration at small distance. Comparison with previous simulation showed significantly better characteristics of the self-injected bunch obtained in this study. The advantages of using Gaussian plasma inhomogeneous profile are demonstrated. An increase in the energy and longitudinal momentum of the bunch in the inhomogeneous case, as well as an increase in the acceleration field in the bunch area were shown.

## ACKNOWLEDGMENTS

The study is supported by the National Research Foundation of Ukraine under the program "Excellent Science in Ukraine" (project # 2023.03/0182).

**ДЖЕРЕЛО ЕЛЕКТРОНІВ НА ОСНОВІ ВИКОРИСТАННЯ САМОІНЖЕКТОВАНОГО ЕЛЕКТРОННОГО ЗГУСТКУ ПРИ ЗБУДЖЕННІ КІЛЬВАТЕРНОГО ПОЛЯ В ПЛАЗМІ ТЕРАВАТНИМ ЛАЗЕРНИМ ІМПУЛЬСОМ**

*Д. С. Бондар, В. І. Маслов, І. М. Оніщенко*

Методи кільватерного прискорення відомі завдяки деяким своїм перевагам. Основним з них є градієнт прискорення до кількох теравольт на метр. Ще однією важливою перевагою, представленою в роботі є можливість використання кільватерного прискорювача як джерела електронів шляхом отримання самоінжектованих згустків та їх прискорення. Результатом є моделювання процесу збудження кільватерного поля плазми лазерним імпульсом з енергією в десятки міліджоулів і потужністю 1-2 ТВт для отримання перспективного джерела електронів. Однорідний та гауссовий профілі плазми були досліджені та порівняні для демонстрації збільшення енергії самоінжектованих згустків. Розглядалися параметри лазера, що відповідають параметрам лазерної установки Інституту плазмової електроніки та нових методів прискорення ННЦ «Харківський фізико-технічний інститут». За результатами моделювання продемонстровано можливість отримання релятивістських самоінжектованих згустків, які можна використовувати для подальших експериментів з лазерним прискоренням, зокрема з діелектричним лазерним прискоренням.